\documentclass[preprint2]{aastex}

\begin{document}

\title{GRBs Light Curves - Another Clue on the Inner Engine}

\author{E. Nakar \& T. Piran}
\affil{Racah Institute for Physics, The Hebrew University, Jerusalem 91904, ISRAEL}

\begin{abstract}
The nature of  the `inner engine' that accelerate and collimate
the relativistic flow at the cores of GRBs is the most
interesting current puzzle concerning GRBs. Numerical simulations
have shown that the internal shocks' light curve reflects the
activity of this inner engine. Using a simple analytic toy model
we clarify the relations between the observed $ \gamma $-rays
light curve and the inner engine's activity and the dependence of
the  light curves on the inner engine's parameters. This simple
model also explains the observed similarity between the observed
distributions of pulses widths and the intervals between pulses
and the correlation between the width of a pulse and the length
of the preceding interval. Our analysis suggests that the
variability in the  wind's Lorentz factors arises due to a
modulation of the mass injected into a constant energy flow.
\end{abstract}

\section{Introduction}

According to the current Fireball model a Gamma-Ray Burst (GRB)
contain four stages: (i) A catastrophic event produces an `inner
pine
 engine'. (ii) This `inner engine' accelerates a barionic wind
into a highly relativistic motion. (iii) Internal collisions
within this wind produce the prompt $ \gamma  $-ray emission (iv)
An external-shock between the wind and the surrounding matter
produces the afterglow. The most mysterious part of this process
is the nature of the `inner engine'. There are no direct
observations of the `inner engine'.

Numerical simulations (Kobayashi, Piran \& Sari 1997; Ramirez \&
Fenimore, 2000) revealed that the $ \gamma  $-rays light curve
replicates the temporal activity of the 'inner engine'. In this
letter we explain, using a simple analytic toy model, these
results. In our toy model the relativistic wind is described as a
sequence of discrete shells with various Lorentz factors, $
\gamma  $. The $ \gamma  $-rays light curve results from
collisions between shells with different values of $ \gamma  $.
We show that the observed time of a pulse (resulting from such a
collision) reflects the time that the inner faster shell was
ejected from the `inner engine'. We show that unless the
background noise prevents the detection of some pulses, the light
curve reflects one third to one half of the shells ejection time.

Nakar \& Piran (2001) discovered that the pulses width, $ \delta
t $, and the intervals between pulses, $ \Delta t $, have a
similar distributions. Moreover, the duration of an interval
between pulses is correlated with the width of the following
pulse. Our analysis shows that in an internal shocks model with
equal energy shells both the intervals and the pulses' widths
reflect the initial separation between the shells.  Therefore,
both observational results arise naturally in this model. If
instead the shells' mass is constant  then the intervals still
reflects the shells' separation but the pulses widths depend also
on the distribution of the shells' Lorentz factors. In this case
the variance in $ \gamma  $ wipes out both the $ \delta t $-$
\Delta t $ similarity and the correlation.

Our toy model includes some simplifying assumptions. We confirm
these results using  numerical simulations.  The equal energy
simulation fit the observations very well, while the equal mass
model does not fit the observations. These results suggest that
the `inner engine' produces a variable Lorentz factor flow by
modulating the mass of a constant energy flow. These results
provide yet another strong support to the Internal Shock model.
They also give one of the first clues on the nature of the `inner
engine'.

\section{The Toy Model\label{Analytical model}}

In our toy model the `inner engine' emits relativistic shells
which collides and produce the observed light curve. We make the
following simplifying assumptions: (i) The shells are  discrete
and homogeneous. Each shell has a well define boundaries and a
well defined $ \gamma $. (ii) The colliding shells merge  into a
single shell after the collision. (iii) Only efficient collisions
produce an observable pulse. The efficiency, $ \varepsilon $, is
defined as the ratio between the post shock internal energy and
the total energy. We consider only collisions with $ \varepsilon
> 0.05$.

Under these assumptions each shell is defined by four parameters
$ t_{i} $, $ m_{i} $, $ \gamma _{i} $ and $ l_{i} $, where $ i $
is the shell index, $ t $ is the ejection time of the shell, and
$ m $, $ \gamma  $, and $ l $ are the mass, Lorentz factor and
width of the shell respectively. For convenience we define $
L_{i,j} $, the interval between the rear end of the i'th shell
and the front of the j'th shell. Note that\footnote{Hereafter we
take c=1. This equality is only approximate since the shells'
velocity is almost (but not exactly) c. } $ L_{i,i+1}\approx
t_{i+1}-(t_{i}+l_{i})$.

\subsection{A single collision}

Consider a single collision between two shells with widths $ l_{1}
$, $ l_{2} $, a separation $ L $, and ejection times $
t_{2}\approx t_{1}+(l_{1}+L) $. We define $ \gamma _{1}\equiv
\gamma  $ and $ \gamma _{2}\equiv a\gamma  $ ($ a>1 $). The
collision efficiency depends strongly on $ a $ (Piran 1999). $
\varepsilon \, (a=2)\approx 0.05 $ , and it decreases fast with
decreasing $ a $. Hence, we consider only collisions with $ a>2 $.

The collision takes place at: $ R_{s}\approx \gamma
^{2}L\frac{2a^{2}}{a^{2}-1}\approx 2\gamma ^{2}L $ ($ R_{s} $ is
measured in the rest frame). Note that as long as $ a>2 $, $
R_{s} $  depends rather weakly on $ \gamma _{2} $. The emitted
photons from the collision reach the observer at time (omitting
the photons flight time):
\begin{equation}
\label{to} t_{obs}\approx t_{1}+l_{1}+R_{s}/(2\gamma ^{2})\approx
t_{1}+l_{1}+L\approx t_{2}
\end{equation}
The photons from the collision are observed almost simultaneously
with an hypothetical photon emitted from the `inner engine'
together with the faster shell (at $ t_{2} $). This result is
accurate up to an error of $ L(a^{2}-1)^{-1} $ , which is small
compared to $ L $ as long as $ a>2 $.

\subsection{Multiple collisions}

\begin{figure}
{\par\centering
\resizebox*{0.95\columnwidth}{!}{\includegraphics{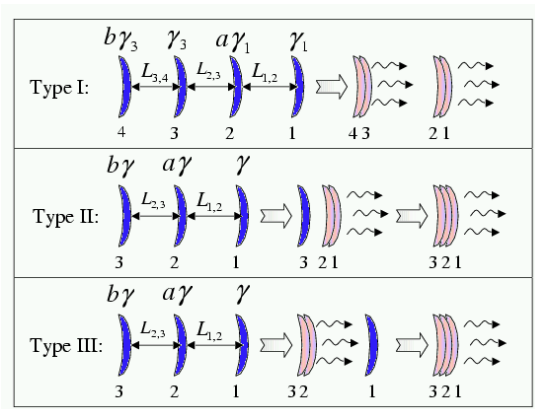}}
\par} \caption{\label{types fig}The basic types of multiple collisions. }
\end{figure}

The light curve obtained from multiple collisions can be
described in terms of three basic pairs of collisions (see fig.
\ref{types fig}): (I) Two collisions between four consequent
shells with $ \gamma _{2}=a\gamma _{1} $ and $ \gamma
_{4}=b\gamma _{3} $. The collision are between the first and the
second shells and between the third and the forth shells. (II)
Two collisions between three consequent shells with $ \gamma
_{1}=\gamma _{2}/a=\gamma _{3}/b $. The two front shells collide
and then the third shell collides with the merged one. (III) Same
as type II but here the rear shells collide first.

\subsubsection{Interval between consequent pulses}

Type I collisions result in two observed pulses, the first at $
t_{2} $ and the second at $ t_{4} $. The interval between the
pulses, $ \Delta t $, would be: $ \Delta t\approx
t_{4}-t_{2}\approx l_{2}+L_{2,3}+l_{3}+L_{3,4} $. In type II
collisions the first pulse is observed at $ t_{2} $, and second
pulse is observed at $ t_{3} $. The interval is: $ \Delta
t\approx t_{3}-t_{2}\approx l_{2}+L_{2,3} $. In type III
collisions the last shell takes over the second one, and the
first pulse is observed at $ t_{3} $. Then the merged shell takes
over the first
one releasing another pulse observed at\footnote{%
Detailed calculations show that the interval between these two
pulses is shorter then the pulses' widths. } $ \sim t_{3} $.
Therefore type III collisions, results in a single wide pulse.

All these results are accurate up to an order of $ 1/a_{i,j}^{2}
$, where $ a_{i,j} $ is the ratio of the Lorentz factors between
the two colliding  shells. As long as the collisions are efficient,
these results  depend weakly on the mass distribution of the
shells.

\subsubsection{The pulses width}

The relevant time scales that determine the pulse width are
(Piran, 1999): (i) The angular time, $ t_{ang} $, which results
from the spherical geometry of the shells: $ t_{ang}\approx
R_{c}/2\gamma ^{2}_{sh} $. (ii) The hydrodynamic time, $ t_{hyd}
$, which arises from the shell's width and the shock crossing
time: $ t_{hyd}\approx l_{c_{in}} $, where $ l_{c_{in}} $ is the
width of the inner shell at the time of the collision.  (iii) The
cooling time - the time that it takes for the emitting electrons
to cool. For synchrotron emission with typical parameters of
internal shocks this time is much shorter then $ t_{ang} $ and $
t_{hyd} $ (Kobayashi et. al. 1997, Wu \& Fenimore 2000). Therefore
under the assumption of transparent shells the pulse width is $
\delta t\approx t_{ang}+t_{hyd} $.

Unlike the pulse's timing, the pulse's width depends strongly on
the shells' masses. The relevant Lorentz factor for the
calculation is the one of the shocked, and therefore radiating,
region - $ \gamma _{sh} $. $ \gamma _{sh} $ depends strongly on
the ratio of the shells' masses. We examine two possible cases:
equal mass shells and equal energy shells. Table \ref{dt-Dt
table} summarizes the intervals and the pulses' width for the two
different mass distributions for  the three types of collisions.

\begin{table*}
{\centering \begin{tabular}{|c|c|c|c|c|c|}
\hline
&
&
\multicolumn{2}{|c|}{{\footnotesize Equal mass}  }&
\multicolumn{2}{|c|}{ {\footnotesize Equal energy}  }\\
\hline & {\footnotesize $ \Delta t $}& {\footnotesize $ \delta
t_{1} $}& {\footnotesize $ \delta t_{2} $}& {\footnotesize $
\delta t_{1} $}&
{\footnotesize $ \delta t_{2} $}\\
\hline {\footnotesize Type I}& {\footnotesize $
l_{2}+L_{2,3}+l_{3}+L_{3,4} $}& {\footnotesize $
l_{c_{2}}+\frac{L_{1,2}}{a} $}& {\footnotesize $
l_{c_{4}}+\frac{L_{3,4}}{b} $}& {\footnotesize $
l_{c_{2}}+L_{1,2} $}&
{\footnotesize $ l_{c_{4}}+L_{3,4} $}\\
\hline {\footnotesize Type II}& {\footnotesize $ l_{2}+L_{2,3}
$}& {\footnotesize $ l_{c_{2}}+\frac{L_{1,2}}{a} $}&
{\footnotesize $
l_{c_{3}}+\frac{L_{1,2}}{b\sqrt{a}}+\frac{L_{2,3}\sqrt{a}}{b} $}&
{\footnotesize $ l_{c_{2}}+L_{1,2} $}&
{\footnotesize $ l_{c_{3}+}\frac{\sqrt{2}}{5}L_{1,2}+L_{2,3} $}\\
\hline {\footnotesize Type III}& {\footnotesize $
\frac{L'_{1,2}}{(ab-1)} $}& {\footnotesize $
l_{c_{3}}+\frac{L_{2,3}}{a} $}& {\footnotesize $
>a\sqrt{\frac{a}{b}}L_{2,3}+\frac{L'_{1,2}\sqrt{ab}}{(ab-1)} $}&
\multicolumn{2}{|c|}{ {\footnotesize No efficient collisions }}\\
\hline
\end{tabular}\footnotesize \par}

\caption{\label{dt-Dt table}  \protect$ \Delta t\protect $, and
\protect$ \delta t\protect $ for the three collisions types for
the equal energy/mass shells. In each case there are two pulses.
\protect$ \delta t_{1}\protect $ {[}\protect$ \delta
t_{2}\protect ${]} is the width of the first {[}second{]}
observed pulse. \protect$ l_{i}\protect $ and \protect$
l_{c_{i}}\protect $ are the width at the ejection and width  at
the collision of the i'th shell  ($l_i=l_{ci}$ with no
spreading.). \protect$ L'_{1,2}\protect $ is the separation
between the first and the second shells at the time that the
second and the third shells collide. The approximation here are
valid when the Lorentz factors ratio between the colliding shells
is larger then 2. }
\end{table*}

\begin{figure}{t}
{\par\centering \resizebox*{0.9\columnwidth}{0.21\textheight}
{\includegraphics{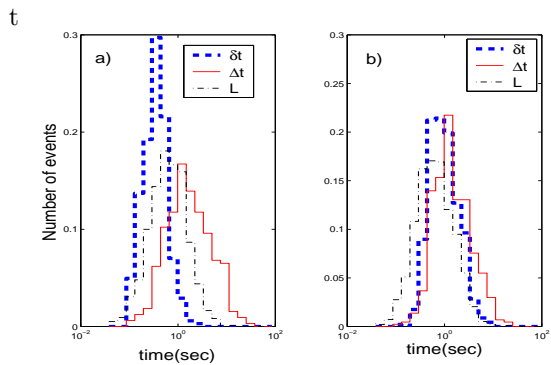}} \par} \caption{\label{dist fig} Pulses
width, \protect$ \delta t\protect $, intervals between pulses,
\protect$ \Delta t\protect $, and the separation between shells,
$L$. (a) : Equal mass shells. (b): Equal energy shells. }
\end{figure}

\section{Numerical Simulation}

The toy model demonstrates that the properties of the light curve
depend on the dominant type of collisions. In order to determine
what are the dominant collisions types we performed numerical
simulations of internal shocks light curves. These simulations
also enable us to verify  some of the approximations used in the
analytic toy model.

 Following Kobayashi et. al. (1997) each
shell  is defined by four parameters $ \gamma _{i},\, m_{i},\,
l_{i} $ and $ L_{i,i+1} $. The distribution of the separation
between the shells, $ L $, is taken to be  lognormal with $ \mu (ln(L))
=-0.5 $ and $ \sigma (ln(L)) =0.9 $ (chosen in order to fit the
observations). The initial shells width is taken as a constant of
0.1lsec and we assume that the shells do not spread ($ l_{c}=l $).
The Lorentz factor distribution is uniform ($ \gamma _{min}=30 $,
$ \gamma _{max}=2000 $). The shells' mass is either constant
(equal mass model) or proportional to $ \gamma ^{-1} $ (equal
energy model).

Each simulation included 50 shells. Following the shell's motion
we identify the collisions. Each collision produce a pulse. The
duration of a pulse is taken as $ t_{ang}+t_{hyd} $. All the
pulses has a fast rise slow decay shape with a ratio of 3:1
between the decay and the rise times. The area below a pulse
equals to its radiated energy (no assumption is made here about
the efficiency). Using these pulses we prepare a binned (64ms
time bins) light curve. We analyze this light curve using the Li
\& Fenimore (1996) peak finding algorithm, obtaining the observed
pulses timing and width.

\subsection{Numerical Results}

In both models the number of observed pulses is between a third
and a half of the total shells emitted. Types I \& II collisions
are the dominant types (about 80\%) in the simulations. The total
efficiency in  both models is about 20-30\%. The equal mass model
prefers  type I collisions while the equal energy model prefer
type II collisions. Efficient type III collisions almost don't
exist in the equal energy model.

Figure \ref{dist fig} illustrates the histograms of the pulses
width, $ \delta t $, the interval between pulses, $ \Delta t $,
and the separation between shells, $ L $. In the equal mass model
(fig. \ref{dist fig}a) there is no $ \Delta t $-$ \delta t $
similarity. $ \Delta t $ reflects the $ L $ distribution while $
\delta t $ is much shorter and does not reflects $ L $. In the
equal energy model (fig. \ref{dist fig}b) both distributions of $
\Delta t $ and $ \delta t $ reflects the `inner engine' shells
separation distribution. Both distributions are consistent with a
log normal distribution and the best fits parameters are
described in table \ref{sim_res_tbl}.
\begin{table}
{\centering \begin{tabular}{|c|c|c|c|c|} \hline &
\multicolumn{2}{|c|}{\( \Delta t\, (sec) \)}&
\multicolumn{2}{|c|}{\( \delta t\, (sec) \)}\\
\hline & \multicolumn{1}{|c|}{\( \mu  \)}& \( 1\sigma  \)& \(
\mu  \)&
\( 1\sigma  \)\\
\hline Simulated& 1.4& 0.6-3.4& 1&
0.5-2\\
\hline Observed& 1.3& 0.5-3.1& 1&
0.5-2.2\\
\hline
\end{tabular}\par}
\caption{\label{sim_res_tbl}Best fit parameters of the pulses'
widths and the intervals in the equal energy model, compared to
the observed values ( Nakar \& Piran 2001). Note that this fit
was achieved by tuning only $\mu (ln(L))$ and $\sigma(ln(L))$.}
\end{table}

Both results are in a perfect agreement with the analytical
results obtained in section \ref{Analytical model}. The
similarity between the {}``observed{}'' interval distribution in
both models is explained by the weak dependence of the pulses'
timing on the mass distribution. The pulses width in the equal
energy model reflection of the shells initial separation, $ L $,
are explained by the analytical result $ \delta t\propto L $. The
deviation from a lognormal distribution and the short pulses in
the equal mass model are explained by the analytical result $
\delta t\propto L/a $.

Nakar \& Piran (2001) find a correlation between an interval
duration and the following pulse. In the equal energy model we
find a highly significant correlation between the interval
duration and the following pulse. There is no significant
correlation in the equal mass model. This result is explained
again by the equal mass relation $ \delta t\propto L/a $. As the
variations in $ a $ are larger then the variations in $ L $ it
wipe out the correlation.

\section{Discussion}

Former numerical simulations of internal shocks have shown that
the resulting light curves reflect the activity of the inner
engine. We have shown that this feature arise from the fact that
the pulse timing is approximately equal to the ejection time of
one of the colliding shell. Moreover, in most collisions (Types I
\& II) the pulses are distinguishable and each pulse reflects a
single collision. In all models the number of observed pulses is
30\%-50\% of the number of ejected shells. Therefore the light
curve reflects the emission time of one third to one half of the
shells. The `inner engine' is slightly more variable then the
observed light curve.

The observed similarity between the $ \Delta t $ and $ \delta t $
distributions is explained naturally in the equal energy shells'
model. Both parameters reflects the the separation between the
shells during their ejection. In the equal mass shells' model
only $ \Delta t $ reflects the initial shells' separation and
therefore such a similarity is not expected.

Our numerical simulations confirmed these predictions.  Note that
many of the simplifying assumptions can be relaxed with no
significant change in the results. We present elsewhere a  more
detailed model and more elaborated simulations (Nakar \& Piran,
2002). The equal energy model simulations fit the observations
very well. These results imply that the `inner engine' ejects,
most likely, equal energy shells. These results are yet another
strong support to the Internal Shock model. They also give one of
the first clues on the nature of the `inner engine'.

This research was supported by a US-Israel BSF grant.


\begin{thebibliography}{1}
\bibitem{1}Kobayashi S., Piran T. \& Sari R., 1997, ApJ, 490, 92
\bibitem{2}Li H. \& Fenimore E., 1996, ApJ, 469, L115
\bibitem{3}Nakar, E. \& Piran, T., 2001, MNRAS in press (astro-ph/0103210)
\bibitem{4}Nakar E. \& Piran, T., 2002 in preperation
\bibitem{5}Piran, T., 1999, PhR, 314, 575
\bibitem{6}Ramirez-Ruiz E. \& Fenimore E. E., 2000, ApJ, 539, 712
\bibitem{7}Wu, B., \& Fenimore, E., 2000 ApJ, 535, L29
\end{thebibliography}
\end{document}